\documentclass[11pt,letterpaper]{article}
\usepackage[margin=1in]{geometry}
\usepackage[T1]{fontenc}
\usepackage{lmodern}
\usepackage{amsmath,amssymb,amsthm}
\usepackage{mathrsfs}
\usepackage{microtype}
\usepackage{graphicx}
\usepackage[nocompress]{cite}
\usepackage[hidelinks]{hyperref}
\hypersetup{pdftitle={Vegetation pattern formation with an energy-mismatch variational closure},pdfauthor={Chad M. Topaz}}
\newcommand{\dd}{\,\mathrm{d}}
\newcommand{\D}{\mathcal D}
\newcommand{\R}{\mathbb R}
\DeclareMathOperator{\Rea}{Re}
\DeclareMathOperator{\Ima}{Im}
\newcommand{\exampleU}{0.1128}
\newcommand{\exampleW}{0.4079}
\newcommand{\weakPeakK}{9.238}
\newcommand{\strongPeakK}{2.145}
\newcommand{\weakGrowth}{0.02307}
\newcommand{\strongGrowth}{0.01545}

\title{Vegetation pattern formation with an\\energy-mismatch variational closure}
\author{Chad M. Topaz\\[3pt]
\small Department of Mathematics and Statistics, Williams College\\
\small Department of Applied Mathematics, University of Colorado Boulder\\
\small \texttt{cmt6@williams.edu}}
\date{September 18, 2026}
\begin{document}
\maketitle
\begin{abstract}
We study a vegetation--water model motivated by a canopy energy mismatch.
Vegetation follows the gradient of a score that rewards biomass and penalizes the squared mismatch, while water obeys a quasi-steady balance.
For a fixed interaction kernel, the linear growth rate about positive uniform vegetation splits into a fixed-water term and a water-feedback term.
The fixed-water term is real, even in wavenumber, and equal to a constant minus a squared modulus, including when the kernel is asymmetric.
All linear phase propagation enters through water feedback.
A second-order expansion of the kernel gives a fourth-order vegetation equation, for which we derive finite-wavenumber growth criteria and corrections from state-dependent spatial coefficients.
The short-wave damping and fastest-growing scales of this truncated model require separate justification as approximations to a specified kernel.
A numerical dispersion example illustrates the growth and phase velocity of its linear modes.
The model creates biomass on bare ground throughout the rainfall range illustrated here, including zero rainfall.
At some smooth nonnegative states, the vegetation growth rate is negative where biomass vanishes.
These features limit its ecological interpretation.

\end{abstract}

\section{Introduction}\label{sec:intro}

Vegetation self-organizes into regular spatial patterns across semi-arid regions worldwide, including the Horn of Africa, the Australian outback, the American Southwest, and the Sahel.
The patterns include bands, often called ``tiger bush,'' as well as spots, labyrinths, and gaps, with wavelengths of tens to hundreds of meters~\cite{Deblauwe2008}.
In a global survey, aridity was the strongest predictor of pattern occurrence, followed by temperature measures and soil texture~\cite{Deblauwe2008}.
Mean annual rainfall alone was a poor predictor.
Because the patterns occur across different plant communities and terrains, modelers have sought general mechanisms that connect plant--environment interactions to spatial organization.

The patterns also evolve in time.
Vegetation bands have been observed to migrate uphill by a few decimeters per year, although other sites show no detectable migration at the available image resolution~\cite{Deblauwe2012}.

Many models use reaction--diffusion--advection equations for vegetation and water.
Klausmeier~\cite{Klausmeier1999} couples plant density and water through a nonlinear growth and uptake term, together with plant mortality, dispersal, and downhill water transport.
Rietkerk and colleagues~\cite{Rietkerk2002} distinguish surface water from soil water and let vegetation enhance infiltration.
Gilad and colleagues~\cite{Gilad2007} include infiltration, shading, and uptake through spatially extended root systems.
In their model, infiltration concentrates soil water beneath vegetation, while root uptake depletes it.
Another approach represents spatial plant interactions through nonlocal cooperation and competition, as in Lefever and Lejeune~\cite{Lefever1997}.

These models differ in their state variables, functional forms, and spatial interactions.
For example, plants in Klausmeier's model take up water where they stand, whereas those in the model of Gilad and colleagues draw water from a neighborhood through their roots.
Here, we ask how a particular growth law divides the linear dynamics into a response at fixed water and feedback through water, and which properties survive changes to the spatial interaction.

Our growth law is motivated by the energy budget of a plant canopy.
Plants shade and shelter their surroundings, while available water limits cooling by transpiration.
We introduce a phenomenological energy mismatch that depends on vegetation density, water, and nearby vegetation.
We then prescribe growth through a score that rewards biomass and penalizes the squared mismatch.
Both the mismatch and its connection to growth are modeling assumptions.

We begin with a polynomial for the local mismatch and an additive linear response to neighboring vegetation through a fixed kernel.
Expanding the kernel integral through second derivatives gives a differential approximation for slowly varying vegetation.
On a hillslope, uphill and downhill neighbors may have different effects, allowing a first-derivative term.
We hold the water field fixed when varying the score and then substitute the solution of a quasi-steady water balance.
This order of operations defines the variational closure, or rule completing the vegetation dynamics.

The resulting dispersion relation splits into a fixed-water term and a water-feedback term.
We first analyze the truncated model, then compare its fixed-water term with the corresponding expression for the full kernel.
The real, even growth rate at fixed water survives that comparison, whereas the quartic damping at short wavelengths belongs to the differential truncation.
We also calculate the effects of state-dependent spatial coefficients and describe the model's failures near bare ground.

\section{Model formulation}\label{sec:model}

We consider a one-dimensional transect of a semi-arid hillslope, with $x$ pointing downhill.
The state variables are vegetation density $u(x,t)$, measured as biomass per unit ground area, and effective water content $w(x,t)$, measured as volume per unit ground area.
The single water field combines plant water availability and spatial transport. Models with separate surface and soil water describe these processes in more detail~\cite{Rietkerk2002,Gilad2007}.
We specify the mismatch and water balance first, then the vegetation growth law in Section~\ref{sec:variational}.

\subsection{Local energy mismatch}\label{subsec:energy}

Leaf and canopy energy budgets depend on radiation, temperature, humidity, wind, and available water~\cite{Campbell1998,Jones2013}.
We describe a local mismatch by a single function $\mathscr G^{(0)}(u,w)$.
Its intended meaning is a proxy for energy input minus dissipation under prescribed environmental conditions.
A physical energy budget would also determine storage and thermal adjustment.
Here, those processes are absent, so the mismatch and the growth rule require their own assumptions.

The local polynomial used here is
\begin{equation}\label{eq:dim_local}
 \mathscr G^{(0)}(u,w)=-\gamma+\alpha_1u-\alpha_2u^2-\beta_1uw,
 \qquad \gamma,\alpha_1,\alpha_2,\beta_1>0.
\end{equation}
The constant $-\gamma$ is an offset.
The term $\alpha_1u$ makes the mismatch rise with density, $-\alpha_2u^2$ reduces its marginal increase, and $-\beta_1uw$ represents water-dependent cooling.
The partial derivatives of $\mathscr G^{(0)}$ are
\begin{equation}\label{eq:dim_derivatives}
 \mathscr G^{(0)}_u=\alpha_1-2\alpha_2u-\beta_1w,
 \qquad \mathscr G^{(0)}_w=-\beta_1u.
\end{equation}
When $\alpha_1>\beta_1w$, the response has a positive-density maximum at $(\alpha_1-\beta_1w)/(2\alpha_2)$.
The cooling contribution vanishes at zero vegetation.

Equation~\eqref{eq:dim_local} is an ansatz for a vegetated regime.
Its concavity, offset, and coefficient signs are choices that would need independent physical justification and calibration.
In particular, the offset has no established canopy interpretation on bare ground.

\subsection{Spatial structure from gradient expansion}\label{subsec:spatial}

The energy budget at one location can depend on vegetation in a surrounding neighborhood through shading, shelter, and other changes to the local environment.
For the fixed-kernel model, we assume an additive linear response to neighboring vegetation, with fixed spatial weights,
\begin{equation}\label{eq:nonlocal}
 \mathscr G[u,w](x)=\mathscr G^{(0)}(u(x),w(x))
 +\eta_d\int_{\R} K(s)\bigl[u(x+s)-u(x)\bigr]\dd s.
\end{equation}
Here, $K$ is a fixed real integrable kernel with units of inverse length, decaying on an interaction length $\ell$, and $\eta_d$ is a constant response amplitude of either sign.
The neighborhood contribution vanishes for uniform vegetation, so the local response remains~\eqref{eq:dim_local}.
Water enters the mismatch through its value at $x$ and obeys the balance in Section~\ref{subsec:water}.
The fixed weights and linear dependence on neighboring density are additional assumptions.
For a nonnegative kernel, positive $\eta_d$ makes denser neighbors increase the mismatch, and negative $\eta_d$ makes them decrease it.
The sign of $\eta_d$ must therefore match whether the proposed effect raises or lowers the mismatch.
Nonlocal plant interactions and their spatial approximations have an established role in vegetation modeling~\cite{Lefever1997}.

For a smooth vegetation field varying on a scale $\lambda\gg\ell$, expanding $u(x+s)$ about $s=0$ gives the moment coefficients
\begin{equation}\label{eq:kernel_moments}
 \ell_1=\int_{\R}sK(s)\dd s,\qquad
 \ell_2=\frac12\int_{\R}s^2K(s)\dd s,
\end{equation}
assuming these moments and the moments needed to control the remainder exist.
Truncating the mismatch at second spatial order gives the truncated model
\begin{equation}\label{eq:dim_G}
 \mathscr G=\mathscr G^{(0)}(u,w)+\eta_d(\ell_1u_x+\ell_2u_{xx}).
\end{equation}
The amplitude $\eta_d$ has units of mismatch per biomass density, while $\ell_1$ and $\ell_2$ have units of length and length squared.
An even kernel gives $\ell_1=0$, as appropriate for a reflection-symmetric flat setting, while a hillslope permits an asymmetric kernel.
Here, the asymmetry is a modeling assumption.
We do not derive its direction or strength from shading or shelter.
Because the kernel term in~\eqref{eq:nonlocal} is linear in $u$, its expansion gives constant coefficients multiplying $u_x$ and $u_{xx}$, with no $u_x^2$ term.
We use~\eqref{eq:dim_G} for the analysis below.
Its validity as an approximation to~\eqref{eq:nonlocal} requires vegetation to vary over distances much larger than the kernel's interaction length.
State-dependent or nonlinear interactions can give additional terms, whose effects we calculate in Section~\ref{subsec:general_spatial}.

\subsection{Water conservation and transport}\label{subsec:water}

The water balance is
\begin{equation}\label{eq:water_cons}
 w_t=R-L(u,w)-J_x,
\end{equation}
where $R\geq0$ is rainfall, $L$ is water loss, and $J$ is the spatial flux.
We use the loss law
\begin{equation}\label{eq:loss}
 L(u,w)=(\mu_0+\mu_1u)w,\qquad \mu_0,\mu_1>0.
\end{equation}
Vegetation can increase transpiration and reduce soil evaporation through shading.
The two effects oppose each other, and we choose $\mu_1>0$ so that the net loss grows with vegetation.
The factor $w$ makes this loss vanish when no water is available.

A first-order gradient expansion of the flux has the form
\begin{equation}\label{eq:flux_general}
 J=J_0(u,w)+J_1(u,w)w_x+J_2(u,w)u_x.
\end{equation}
We choose
\begin{equation}\label{eq:dim_flux}
 J=(v_0-v_1u)w-d_0w_x-\delta_0wu_x,
 \qquad v_0,v_1,\delta_0\geq0,\quad d_0>0.
\end{equation}
The first term represents density-dependent slowing of advective transport, and the second gives diffusive redistribution.
For a downhill-flow interpretation, $v_0-v_1u$ must remain positive in the regime of use.
This flux does not resolve a contrast in infiltration between bare and vegetated ground.
The last term moves water toward sparser vegetation when $\delta_0>0$.
The divergence of the advective term already contains $-v_1wu_x$, whereas the last term contributes $wu_{xx}$ and $w_xu_x$.
These are distinct effects, so we treat $\delta_0$ as an independent, uncalibrated coefficient.

We adopt the quasi-steady approximation
\begin{equation}\label{eq:dim_water}
 0=R-(\mu_0+\mu_1u)w-\partial_x\left[(v_0-v_1u)w-d_0w_x-\delta_0wu_x\right],
\end{equation}
and denote its solution by $w=\mathcal W[u]$ when it is uniquely determined by the boundary conditions.
This assumes that the effective water field adjusts rapidly relative to vegetation.
It removes rainfall-event dynamics and the associated transient redistribution.

Although the flux law is local, water at one point depends on vegetation at other points.
For example, in the simplified advective case $d_0=\delta_0=v_1=0$, with $v_0>0$ and an illustrative upstream boundary condition $w(0)=0$,
\begin{align}
 v_0w_x&=R-(\mu_0+\mu_1u)w,\label{eq:upstream_ode}\\
 w(x)&=\frac{R}{v_0}\int_0^x\exp\left[-\frac{1}{v_0}\int_y^x(\mu_0+\mu_1u(\xi))\dd\xi\right]\dd y.\label{eq:upstream_solution}
\end{align}
Water at $x$ depends on losses along the upstream flow path.
In the stability analysis below, we retain $d_0>0$ and impose periodic boundary conditions on a short stretch of a long transect.

\section{Variational closure and dimensionless model}\label{sec:variational}

We complete the model by prescribing how vegetation responds to the mismatch.
On a periodic interval $\Omega$, define the score
\begin{equation}\label{eq:dim_score}
 \mathscr S[u;w]=\int_\Omega\left(u-\frac{\kappa}{2}\mathscr G^2\right)\dd x,
 \qquad \kappa>0.
\end{equation}
Here, $\kappa$ weights the squared mismatch relative to the biomass reward.
The penalty treats positive and negative mismatches equally.
We let vegetation follow the gradient of this score,
\begin{equation}\label{eq:gradient_law}
 u_t=\tau\left.\frac{\delta\mathscr S}{\delta u}\right|_{w\ {\rm fixed}},\qquad \tau>0,
\end{equation}
where $\tau$ is a rate coefficient.
Taking the functional derivative with the whole function $w(x)$ held fixed gives
\begin{equation}\label{eq:dim_dynamics}
 u_t=\tau\left[1-\kappa\mathscr G\mathscr G^{(0)}_u
       +\kappa\eta_d\ell_1\mathscr G_x
       -\kappa\eta_d\ell_2\mathscr G_{xx}\right]_{w=\mathcal W[u]},\qquad \tau>0.
\end{equation}
Integration by parts on $\Omega$ eliminates the boundary terms.
The second derivative in the mismatch produces fourth-order spatial derivatives in the vegetation equation.

During this variation, water can still vary in space.
Afterward, we substitute the quasi-steady solution $w=\mathcal W[u]$ as in~\eqref{eq:dim_dynamics}.
Varying the composed score $\mathscr S[u;\mathcal W[u]]$ would add terms involving the derivative of $\mathcal W$.
Two features of~\eqref{eq:dim_dynamics} matter near bare ground.
The biomass reward contributes the constant $\tau$, and the right side carries no factor of $u$, so growth need not vanish where vegetation does.
Section~\ref{subsec:scope} describes the resulting limitations.

\subsection{Nondimensionalization}\label{subsec:scaling}

We scale biomass, water, mismatch, and time by
\begin{equation}\label{eq:scales}
 u_* = \frac{\alpha_1}{2\alpha_2},\qquad
 w_* = \frac{\alpha_1}{\beta_1},\qquad
 G_* = \alpha_1u_*,\qquad
 t_* = \frac{u_*^2}{\tau\kappa G_*^2}.
\end{equation}
The scale $u_*$ is the location of the maximum of the water-independent quadratic, and $w_*$ is the water level at which the coefficient $\alpha_1-\beta_1w$ vanishes.
Choose a positive reference length $x_*$ and set $U=u/u_*$, $W=w/w_*$, $X=x/x_*$, and $T=t/t_*$.
On a slope, a natural choice is $x_*=v_0/\mu_0$.
We leave $x_*$ free so that the same scaling covers flat ground, where $v_0=0$.
The dimensionless parameters are
\begin{align}
 \mu&=\frac{u_*}{\kappa G_*^2}, &
 \Gamma&=\frac{\gamma}{G_*}, &
 \rho&=\frac{R}{\mu_0w_*}, &
 \beta&=\frac{\mu_1u_*}{\mu_0},\label{eq:params_local}\\
 \eta&=\frac{\eta_du_*}{G_*}, &
 \Lambda_1&=\frac{\ell_1}{x_*}, &
 \Lambda_2&=\frac{\ell_2}{x_*^2}, &
 \nu&=\frac{v_0}{\mu_0x_*},\label{eq:params_spatial}\\
 \D&=\frac{d_0}{\mu_0x_*^2}, &
 \Delta&=\frac{\delta_0u_*}{\mu_0x_*^2}, &
 \chi&=\frac{v_1u_*}{v_0}\quad(v_0>0).&&\label{eq:params_water}
\end{align}
The parameter $\mu$ measures the biomass reward relative to the mismatch penalty, so small $\mu$ corresponds to a strong relative penalty.
For compactness, write $a=\eta\Lambda_1$ and $b=\eta\Lambda_2$.
We assume $b\ne0$ in the fourth-order analysis.
On flat ground, we set $v_0=v_1=0$.
Then $\nu=0$, and $\chi$ drops out because it enters the flux only through $\nu\chi$.
Reflection symmetry also gives $a=0$.

\subsection{The resulting system}\label{subsec:system}

The dimensionless score has density $\mu U-G^2/2$, with
\begin{equation}\label{eq:G}
 G=-\Gamma+U-\tfrac12U^2-UW+aU_X+bU_{XX}.
\end{equation}
The dimensionless model is
\begin{align}
 U_T&=F[U,W]=\mu-G(1-U-W)+aG_X-bG_{XX},\label{eq:F}\\
 0&=\rho-(1+\beta U)W-J_X,\label{eq:water}\\
 J&=\nu(1-\chi U)W-\D W_X-\Delta WU_X.\label{eq:flux}
\end{align}
The right side of~\eqref{eq:F} is the functional derivative of the dimensionless score with respect to $U$ at fixed $W$.
When $W$ is prescribed and independent of time, that score is nondecreasing along smooth periodic solutions.

\section{Linear stability analysis}\label{sec:lsa}
\subsection{Uniform steady states}

At a uniform equilibrium $(U_0,W_0)$ with $U_0>0$, the water balance fixes $W_0$.
We denote the mismatch by $G_0$ and its partial derivative with respect to $U$ at fixed $W$ by $q$, so that
\begin{equation}\label{eq:uniform}
 W_0=\frac{\rho}{1+\beta U_0},\qquad
 q=1-U_0-W_0,\qquad
 G_0=-\Gamma+U_0-\tfrac12U_0^2-U_0W_0.
\end{equation}
The remaining steady-state equation is
\begin{equation}\label{eq:steady}
 \mu=G_0q.
\end{equation}
This equation can have several positive roots, and the formulas below hold at each of them.
Since $\mu>0$, neither $G_0$ nor $q$ vanishes at such a state.

\subsection{Water response and dispersion relation}

For perturbations proportional to $e^{ikX+\sigma T}$, write $\widehat U$ and $\widehat W$ for their amplitudes.
Linearizing~\eqref{eq:water}--\eqref{eq:flux} gives
\begin{equation}\label{eq:waterlinear}
 \left(1+\beta U_0+\D k^2+i c_0k\right)\widehat W
 =-W_0\left(\beta+\Delta k^2-i\chi\nu k\right)\widehat U,
 \quad c_0=\nu(1-\chi U_0).
\end{equation}
The coefficient of $\widehat W$ has positive real part, so we can divide by it.
The water response is $\widehat W=\Phi(k)\widehat U$, where
\begin{equation}\label{eq:Phi}
 \Phi(k)=-\frac{W_0(\beta+\Delta k^2-i\chi\nu k)}
 {1+\beta U_0+\D k^2+i c_0k}.
\end{equation}

Substituting this water response into the linearized vegetation equation gives the dispersion relation for the model~\eqref{eq:G}--\eqref{eq:flux} at a positive uniform equilibrium,
\begin{equation}\label{eq:dispersion}
 \boxed{\displaystyle
 \sigma(k)=\underbrace{G_0-|q+iak-bk^2|^2}_{H_{\rm fw}(k)}
 +\underbrace{(qU_0+G_0-iaU_0k-bU_0k^2)}_{\mathcal B(k)}\Phi(k).}
\end{equation}
To derive this expression, set $\sigma\widehat U=\widehat{\delta F}$.
The multiplier of $\widehat U$ in the linearized mismatch is $P=q+iak-bk^2$.
The multiplier of the operator acting on $G$ in~\eqref{eq:F} is $M=-q+iak+bk^2=-\overline P$.
Thus
\begin{equation}\label{eq:linearG}
 \widehat{\delta G}=P\widehat U-U_0\widehat W,
 \qquad \widehat{\delta(1-U-W)}=-\widehat U-\widehat W.
\end{equation}
Linearizing~\eqref{eq:F} therefore gives
\begin{equation}\label{eq:linearF}
 \widehat{\delta F}=(MP+G_0)\widehat U+(-U_0M+G_0)\widehat W.
\end{equation}
Using $MP=-|P|^2$ and substituting~\eqref{eq:Phi} gives~\eqref{eq:dispersion}.

The fixed-water term $H_{\rm fw}$ is the growth rate when water is held at $W_0$.
The water-feedback term $\mathcal B\Phi$ is the change caused by the response of water.
The coefficient multiplying $\Phi$ can also be written as $\mathcal B=G_0+U_0\overline P$.

\subsection{Fixed-water growth and finite-wavenumber instability}

Expanding the first term in~\eqref{eq:dispersion} gives
\begin{equation}\label{eq:local}
 H_{\rm fw}(k)=G_0-q^2+(2bq-a^2)k^2-b^2k^4\leq G_0.
\end{equation}
It is real and even in $k$, and it can be positive only if $G_0>0$.
If $G_0\leq0$, any instability must come from water feedback.

Let $\theta=2bq-a^2$ denote the coefficient of $k^2$.
Suppose $H_{\rm fw}(0)=G_0-q^2<0$.
Then $H_{\rm fw}(k)>0$ for some real $k\ne0$ if and only if
\begin{equation}\label{eq:criterion}
 \theta>0,\qquad G_0-q^2+\frac{\theta^2}{4b^2}>0.
\end{equation}
Maximizing the quadratic in $k^2$ gives $k^2=\theta/(2b^2)$ when $\theta>0$.
Otherwise, the maximum occurs at zero.

When $a=0$, $b>0$, and $q>0$, the fixed-water maximum is $G_0=\mu/q>0$ at $k^2=q/b$.
Every such state therefore has a positive fixed-water maximum in this truncated model.
At the same wavenumber, $P=0$ and the full growth rate is
\begin{equation}\label{eq:coupled_sufficient}
 \sigma\bigl(\sqrt{q/b}\bigr)
 =G_0\left[1+\Phi\bigl(\sqrt{q/b}\bigr)\right].
\end{equation}
Since $G_0>0$, the condition $\Rea\Phi(\sqrt{q/b})>-1$ is sufficient for positive coupled growth there.

On a periodic interval of length $\mathcal L$, the allowed wavenumbers are $k\in(2\pi/\mathcal L)\mathbb Z$.
The maximizer over real $k$ may fall between them.
We call the coupled state unstable at finite wavenumber when $\sigma(0)<0$ and $\Rea\sigma(k)>0$ at some allowed $k\ne0$.

For $b\ne0$ and $\D>0$, the water response $\Phi(k)$ is bounded at large $|k|$, while $\mathcal B(k)=O(k^2)$.
More precisely,
\begin{equation}\label{eq:highk}
 \Rea\sigma(k)=-b^2k^4+\left(2bq-a^2+\frac{bU_0W_0\Delta}{\D}\right)k^2+O(1).
\end{equation}
Thus the truncated model damps sufficiently short waves.
This large-wavenumber expansion need not be uniform as $\D\to0$.
With $a=\Delta=0$, the $O(1)$ term approaches $G_0-q^2+bU_0W_0(\beta\D-\chi\nu c_0)/\D^2$, whose magnitude diverges as $\D\to0$ when $W_0>0$.
In the formal limiting case $\D=\nu=0$ with $\Delta>0$, the quartic coefficient itself becomes
$-b^2+bU_0W_0\Delta/(1+\beta U_0)$ and can be positive.

\subsection{Comparison with the full interaction kernel}\label{subsec:full_kernel}

The negative-square form survives without a gradient truncation.
Let $\mathcal K(S)=x_*K(x_*S)$ be the dimensionless kernel and use the Fourier convention
$\widehat{\mathcal K}(k)=\int_{\R}\mathcal K(S)e^{ikS}\dd S$.
For the full mismatch~\eqref{eq:nonlocal}, its linear multiplier is
\begin{equation}\label{eq:kernel_symbol}
 P_K(k)=q+\eta\bigl[\widehat{\mathcal K}(k)-\widehat{\mathcal K}(0)\bigr].
\end{equation}
With the full mismatch, variation moves the kernel onto $G$ with its argument reflected,
\begin{equation}\label{eq:kernel_growth_law}
 U_T=\mu-G(1-U-W)
 -\eta\int_{\R}\mathcal K(S)\bigl[G(X-S)-G(X)\bigr]\dd S.
\end{equation}
On a Fourier mode, the integral acting on $G$ has the complex conjugate of the multiplier of the integral acting on $U$.
Linearization therefore gives
\begin{equation}\label{eq:kernel_dispersion}
 H_{{\rm fw},K}(k)=G_0-|P_K(k)|^2,\qquad
 \sigma_K(k)=H_{{\rm fw},K}(k)+\bigl(G_0+U_0\overline{P_K(k)}\bigr)\Phi(k).
\end{equation}
Because $\mathcal K$ is real, $P_K(-k)=\overline{P_K(k)}$.
Hence the fixed-water term is real, even, and bounded above by $G_0$, including for an asymmetric kernel.
Kernel asymmetry alone cannot make linear modes travel in this variational model with water fixed.

Integrability implies $\widehat{\mathcal K}(k)\to0$ as $|k|\to\infty$, and therefore
\begin{equation}\label{eq:kernel_highk}
 H_{{\rm fw},K}(k)\longrightarrow
 G_0-\left[q-\eta\int_{\R}\mathcal K(S)\dd S\right]^2.
\end{equation}
This limit can have either sign.
For $\D>0$, the full growth rate also approaches a finite real limit, obtained from~\eqref{eq:kernel_dispersion} using $\Phi\to-W_0\Delta/\D$.
The negative quartic term in~\eqref{eq:highk} thus comes from truncating the mismatch at second derivatives.

For a centered Gaussian kernel, the wavenumber at which the fixed-water term is largest can be compared with the truncated value.
Assume $q>0$, and let $N=\eta\int_{\R}\mathcal K(S)\dd S>0$ be the total interaction strength and $\ell_K>0$ the dimensionless standard deviation.
Then $b=N\ell_K^2/2$, and $P_K=q+N(e^{-\ell_K^2k^2/2}-1)$.
If $N>q$, the full fixed-water term reaches $G_0$ at
\begin{equation}\label{eq:gaussian_peak}
 k_K^2=-\frac{N}{b}\log(1-q/N),\qquad
 \frac{k_K}{\sqrt{q/b}}=1+\frac{q}{4N}+O\bigl((q/N)^2\bigr).
\end{equation}
Thus the truncated wavenumber is accurate in this Gaussian family when $q/N$ is small.
At the state used in Section~\ref{subsec:example} with $b=0.005$, taking $\ell_K=0.1$ gives $N=1$ and $q/N\approx0.48$, and the exact wavenumber $k_K\approx11.42$ exceeds the truncated value $\sqrt{q/b}\approx9.790$ by about seventeen percent.
At that scale $k_K\ell_K\approx1.1$, so this example lies outside the long-wavelength regime assumed in Section~\ref{subsec:spatial}.
For $0<N\leq q$, the full fixed-water term increases with $|k|$ and approaches its supremum only at infinity.
Its limiting sign is determined by~\eqref{eq:kernel_highk}.
A physical use of the predicted scale would require estimating the kernel strength and width.

\subsection{Water feedback and linear phase propagation}

Let $A(k)=1+\beta U_0+\D k^2$ denote the real part of the denominator of $\Phi$.
Equation~\eqref{eq:Phi} gives
\begin{equation}\label{eq:RePhi}
 \Rea\Phi(k)=-\frac{W_0[(\beta+\Delta k^2)A(k)-\chi\nu c_0k^2]}{A(k)^2+c_0^2k^2}.
\end{equation}
The loss and gradient-flux terms contribute negatively to $\Rea\Phi$.
When $c_0>0$ and $\chi\nu>0$, the advective term acts in the opposite direction and can outweigh them.
The growth rate contains $\Rea(\mathcal B\Phi)$, however, so the sign of $\Rea\Phi$ alone does not settle the effect on growth.
The coefficient $\Delta$ changes the water response at nonzero wavenumber while leaving the uniform equilibrium fixed.

All linear phase propagation enters through
\begin{equation}\label{eq:imaginary}
 \Ima\sigma(k)=\Ima[\mathcal B(k)\Phi(k)].
\end{equation}
A linear mode with $k\ne0$ has phase velocity $c_{\rm ph}=-\Ima\sigma(k)/k$.
Because $X$ increases downhill, negative velocity means uphill motion.
The asymmetry coefficient $a$ appears in $\mathcal B$, so it can affect phase propagation even when $\Phi$ is real.
For example, if $\nu=0$, then $\Ima\sigma=-aU_0k\Phi$.
For $\nu=a=0$, both factors are real and the linear modes have zero phase velocity.

\subsection{An illustrative dispersion calculation}\label{subsec:example}

\begin{figure}[p]
\centering
\includegraphics[width=\textwidth]{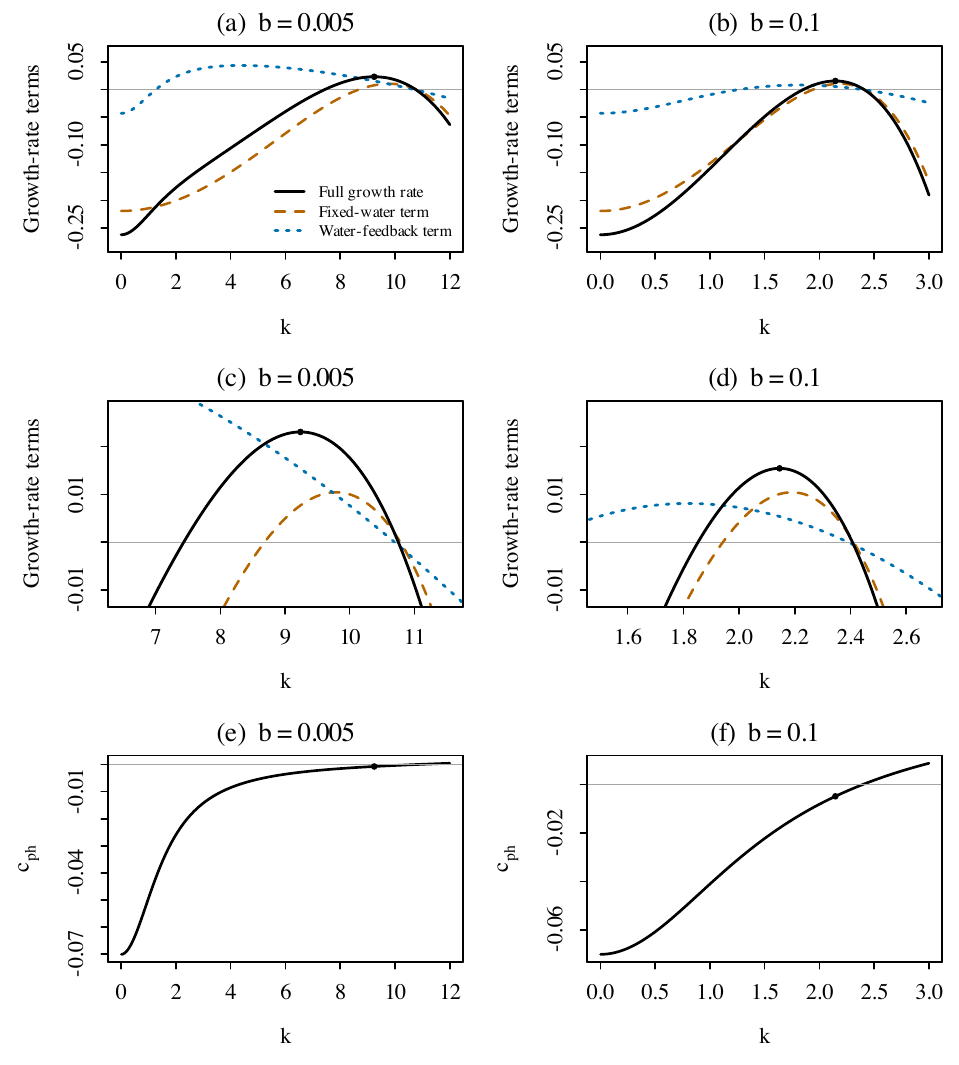}
\caption{Linear growth and phase velocity for the example in Section~\ref{subsec:example}.
The left column has $b=0.005$ and the right column $b=0.1$, with different horizontal scales.
The upper row shows the full growth rate and its two terms, and the middle row enlarges the region near the maxima.
The solid black curve is $\Rea\sigma$, the dashed orange curve is $H_{\rm fw}$, and the dotted blue curve is $\Rea(\mathcal B\Phi)$.
The dashed and dotted curves sum to the solid curve, whose maxima are marked by dots.
The lower row shows $c_{\rm ph}$, with negative values corresponding to uphill phase propagation and dots marking the fastest-growing wavenumbers.
All other parameters are identical across the columns.}\label{fig:dispersion}
\end{figure}

Figure~\ref{fig:dispersion} shows the dispersion relation for two values of $b$.
We set $\Gamma=0.05$, $\mu=0.005$, $\rho=0.5$, $\beta=2$, $\D=0.01$, $\chi=2$, $\nu=1$, and $a=\Delta=0$, and compare $b=0.005$ with $b=0.1$.
For a nonnegative kernel, these positive coefficients correspond to denser neighbors raising the mismatch.
The uniform equation has three positive roots, approximately $0.1128$, $0.7876$, and $1.7241$.
We use the smallest, $U_0=\exampleU$ and $W_0=\exampleW$, which is the only one with $1-\chi U_0>0$.
It is stable to uniform perturbations, with $\sigma(0)\approx-0.2622$, and unstable on an interval of nonzero wavenumbers in both cases.

Since $a=0$, the two fixed-water curves differ only by rescaling $k$ by $\sqrt b$ and have the same maximum $G_0\approx0.01043$.
Changing $b$ also changes $\mathcal B$ and hence the water-feedback term.
The fastest-growing wavenumber shifts from $k_{\max}=\weakPeakK$ to $k_{\max}=\strongPeakK$, and the maximum growth rate falls from $\weakGrowth$ to $\strongGrowth$.
At these two maxima, water feedback is positive, contributing approximately $0.01539$ and $0.005373$, respectively.
It supplies about two thirds of the peak growth in the first case.
The example has a symmetric mismatch interaction and downhill water advection, with uphill phase propagation at both maxima.
Since $a=0$, the real factor $\mathcal B$ vanishes at $bk^2=q+G_0/U_0$, giving $k\approx10.69$ and $2.391$.
Water feedback and phase velocity change sign there, so the narrow upper ends of the unstable intervals contain modes propagating downhill.

\subsection{Evenness and the choice of growth law}

An even fixed-water growth rate is a consequence of the variational construction.
We use evenness to determine relations among the coefficients of growth laws built from $G$, $G_X$, and $G_{XX}$.

The linearized mismatch has Fourier multiplier $P$, and its adjoint has multiplier $\overline P$.
Linearizing the variational growth law produces the term $-\overline P P=-|P|^2$, which is real and even because $P(-k)=\overline{P(k)}$.
The additional fixed-water term $G_0$ in~\eqref{eq:dispersion} comes from the second derivative of the local mismatch with respect to $U$.
More generally, the second variation of a real functional is self-adjoint; at a uniform state, translation invariance then gives a real, even Fourier multiplier.

Consider growth laws of the form
\begin{equation}\label{eq:closure_family}
 U_T=f(U,W)+m(U,W)G+n(U,W)G_X-r(U,W)G_{XX},
\end{equation}
with smooth real functions $f,m,n,r$ and the mismatch~\eqref{eq:G}.
At a uniform equilibrium, the fixed-water growth rate is
\begin{equation}\label{eq:family}
 (m_0+in_0k+r_0k^2)(q+iak-bk^2)+f_U+G_0m_U,
\end{equation}
where subscripts $0$ and all derivatives denote evaluation at the equilibrium.
For $a\ne0$, setting the coefficients of $k$ and $k^3$ to zero gives
\begin{equation}\label{eq:evenness}
 m_0a+n_0q=0,\qquad -n_0b+r_0a=0.
\end{equation}
Solving these equations gives $(m_0,n_0,r_0)=\zeta(-q,a,b)$ for some real $\zeta$; the variational law~\eqref{eq:F} has $\zeta=1$.
Thus evenness fixes the relative values of these three coefficients at the chosen equilibrium.
Their common factor has arbitrary sign, while a strictly negative quartic coefficient $-\zeta b^2$ requires $b\ne0$ and $\zeta>0$.
Many nonlinear growth laws can share these coefficient values at one equilibrium.

For the symmetric mismatch $a=0$ with $b\ne0$, the only odd powers of $k$ come from the $n_0$ term in~\eqref{eq:family}.
Evenness then requires $n_0=0$ and leaves $m_0$ and $r_0$ free.

\subsection{State-dependent spatial coefficients}\label{subsec:general_spatial}

We now let the spatial coefficients depend on the state, while keeping the local mismatch and water equation.
For smooth real functions $h,j,p$, take
\begin{equation}\label{eq:general_G}
 G=g(U,W)+h(U,W)U_X+j(U,W)U_{XX}+p(U,W)U_X^2,
 \qquad g=-\Gamma+U-\tfrac12U^2-UW.
\end{equation}
Taking the functional derivative of the score at fixed $W$ gives
\begin{equation}\label{eq:general_F}
 \begin{split}
 F={}&\mu-G\bigl(g_U+h_UU_X+j_UU_{XX}+p_UU_X^2\bigr)\\
     &+\partial_X\bigl[G(h+2pU_X)\bigr]-\partial_X^2(Gj).
 \end{split}
\end{equation}

The uniform-state equations and water response $\Phi(k)$ are unchanged.
We reuse $a,b$ for $h(U_0,W_0),j(U_0,W_0)$ and write $p_0=p(U_0,W_0)$.
All coefficient derivatives below are evaluated at $(U_0,W_0)$.
Let $\sigma_{\rm fr}(k)$ denote~\eqref{eq:dispersion} evaluated with these frozen values $a,b$.
Linearizing~\eqref{eq:general_F} yields
\begin{equation}\label{eq:general_dispersion}
 \boxed{\displaystyle
 \sigma_{\rm gen}(k)=\sigma_{\rm fr}(k)
 +2G_0(j_U-p_0)k^2
 +G_0\bigl(ih_Wk+j_Wk^2\bigr)\Phi(k).}
\end{equation}

The linearization of $G$ is still
$(q+iak-bk^2)\widehat U-U_0\widehat W$.
The term $-Gh_UU_X$ and the $h_U$ part of $\partial_X(Gh)$ cancel at linear order.
The remaining coefficient derivatives give the two corrections in~\eqref{eq:general_dispersion}.
Derivatives of $p$ do not enter at a uniform state.
When $h,j,p$ are constants, the only correction is $-2p_0G_0k^2$.
Although $pU_X^2$ contributes to $G$ only at second order in the perturbation of $U$, the variation multiplies it by the nonzero mismatch $G_0$, so it affects the linear evolution.

The fixed-water term remains real and even, with quadratic coefficient
\begin{equation}\label{eq:general_d}
 \theta_{\rm gen}=2bq-a^2+2G_0(j_U-p_0).
\end{equation}
When $H_{\rm fw}(0)<0$ and $b\ne0$, the criterion~\eqref{eq:criterion} applies with $\theta$ replaced by $\theta_{\rm gen}$.
For $b\ne0$ and $\D>0$, this generalized truncated model still has $\Rea\sigma=-b^2k^4+O(k^2)$ at large $|k|$.
As before, only the water-feedback term can be complex.
The bound $H_{\rm fw}\leq G_0$ and the specific instability thresholds of the constant-coefficient truncated model need not persist.

\section{Discussion}\label{sec:discussion}

\subsection{Relationship to existing models}

The models of Klausmeier, Rietkerk and colleagues, and Gilad and colleagues connect plant growth with water redistribution through different assumptions~\cite{Klausmeier1999,Rietkerk2002,Gilad2007}.
Klausmeier uses a single water field and local uptake.
Rietkerk and colleagues separate surface and soil water, while Gilad and colleagues also represent spatially extended roots and canopy-dependent evaporation.
Our model relates coefficients in the vegetation equation through a mismatch penalty and variation at fixed water.
Its spatial vegetation terms arise from that construction; no separate dispersal term is included.
At fixed water, plant diffusion in the Klausmeier model contributes a negative multiple of $k^2$ to the growth rate.
Here, the differential truncation gives $-b^2k^4$ and a coefficient $2bq-a^2$ of $k^2$ that can have either sign.

Lefever and Lejeune~\cite{Lefever1997} already derive differential approximations from nonlocal plant interactions.
Their Appendix B retains reproduction and dispersal terms through second derivatives and inhibition through fourth derivatives to control short waves.
They also obtain traveling patterns by shifting an interaction kernel, which gives an imaginary part in the linear growth rate.
In our model, the kernel and its adjoint instead enter together through $|P_K|^2$.
The resulting fixed-water term is real even for an asymmetric kernel, and phase propagation enters through the product $\mathcal B\Phi$.

For $a=0$ and $q/b>0$, the truncated fixed-water term can be written as
\begin{equation}\label{eq:sh_form}
 H_{\rm fw}(k)=G_0-b^2(k^2-q/b)^2.
\end{equation}
This is the familiar linear form of a Swift--Hohenberg operator~\cite{SwiftHohenberg1977}, with critical wavenumber $\sqrt{q/b}$.
The squared-penalty construction provides one route to that form.
As Section~\ref{subsec:full_kernel} shows, extending its short-wave damping to the full interaction law would require further assumptions on the kernel and parameters.

\subsection{Scope of the model}\label{subsec:scope}

We have not calibrated the coefficients to any ecosystem.
Calibration would require measurable definitions of the mismatch and effective water content, together with tests of the gradient expansion and quasi-steady approximation.
Without calibrated length and time scales, the dimensionless results in Figure~\ref{fig:dispersion} cannot be compared quantitatively with observed wavelengths or migration speeds.
Water enters the mismatch as a coolant, while the vegetation equation has a constant biomass reward.
These choices fail to enforce water limitation of growth.

On uniform bare ground, $U=0$, $W=\rho$, and
\begin{equation}\label{eq:bare}
 F[0,\rho]=\mu+\Gamma(1-\rho).
\end{equation}
Bare ground is an equilibrium only at $\rho=1+\mu/\Gamma$.
For every smaller rainfall, including zero, the model creates biomass on bare ground.
For the example parameters, this occurs throughout $0\leq\rho<1.1$.
For larger rainfall, the same formula gives negative growth at zero biomass.
Appendix~\ref{app:positivity} constructs smooth nonnegative vegetation data at $\rho=0.5$ for which the spatial terms make the growth rate negative at a zero of biomass.

At $\rho=0$, the uniform water level vanishes, but the vegetation equation still has three positive equilibria, approximately $U_0=0.05692$, $0.9889$, and $1.954$ for $\Gamma=0.05$ and $\mu=0.005$.
The smallest is stable to uniform perturbations and, when $a=0$ and $b>0$, has a band of positive growth in the continuous dispersion relation with no water feedback.
A periodic interval is unstable precisely when it admits a wavenumber in that band.
This zero-rainfall growth further limits the ecological interpretation.

For the parameters of Section~\ref{subsec:example} with $b=0.005$, following the sparse uniform branch numerically gives a fold near $\rho=0.9353$.
Its maximum linear growth rate over real wavenumbers rises from about $0.0053$ at zero rainfall to about $0.0892$ near $\rho=0.927$, before declining toward the fold.
The dimensionless wavelength of its fastest-growing linear mode increases from about $0.458$ at zero rainfall to about $1.350$ near $\rho=0.91$, then decreases to about $1.304$ just below the fold.

The dense equilibrium is also uniformly stable.
Following the dense branch with $\beta=2$ decreases its biomass to approximately $1.380$ at $\rho=1.05$.
The dense equilibria lie beyond the scale $U=1$ of the local mismatch maximum at zero water, so they are formal solutions of the polynomial model.

For $\nu>0$ and $\chi>0$, the density-dependent advection speed is downhill only while $U<1/\chi$.
Its reversal above that density marks another limit of the flux approximation.
Thus the dense branches of the illustrative polynomial system lie outside the intended downhill-flow regime when $\chi=2$.
For nonuniform vegetation, the coefficient of $W$ in the flux also contains the gradient term $-\Delta U_X$.

These results are linear growth rates about uniform equilibria.
Whether the model sustains nonlinear patterns, and with what wavelength and migration speed, is a separate question that we do not address.
A nonlinear ecological study would first require justified growth and transport laws that respect biomass nonnegativity and water limitation, followed by a verified treatment of the coupled spatial equations.

\section{Conclusion}\label{sec:conclusion}

For the fixed-kernel variational law, the fixed-water growth rate is real, even, and bounded above by the uniform mismatch.
These properties survive removal of the gradient truncation, including for asymmetric kernels.
The second-order expansion of the mismatch gives a fourth-order vegetation equation whose finite-wavenumber growth and water feedback can be calculated explicitly.
For $b\ne0$ and $\D>0$, the truncated growth rate tends to $-\infty$ quartically, whereas the full-kernel rate approaches a finite limit that can have either sign.
The failures near bare ground and at zero rainfall leave the model's ecological justification unresolved.

\section*{Acknowledgments and disclosure}
This work received no external funding.
Claude (Anthropic) and ChatGPT (OpenAI) assisted with editing, consistency checks, and prose. Responsibility for the manuscript and results rests with the author.

\appendix
\section{A counterexample to biomass nonnegativity}\label{app:positivity}

We construct smooth nonnegative initial data for which~\eqref{eq:F}--\eqref{eq:flux} give a negative vegetation growth rate at a zero of $U$.
On a flat periodic interval of length $2\pi/k$, take $a=\nu=0$ and
\begin{equation}\label{eq:counterexample}
 U(X)=A_{\rm b}(1-\cos kX)^2,\quad A_{\rm b}=0.01,\quad k=10,
 \quad \Gamma=0.05,\quad\mu=0.005,\quad b=0.01,
\end{equation}
with $\rho=0.5$, $\beta=2$, $\D=0.01$, and $\Delta=0.1$.
At $X=0$, $U$ and its first three derivatives vanish, while $U_{XXXX}=6A_{\rm b}k^4=600$.
At $X=0$, we therefore have $G=-\Gamma$ and $G_{XX}=bU_{XXXX}=6$, so that
\begin{equation}\label{eq:outward}
 F[U,W](0)=0.005+0.05(1-W(0))-0.06=-0.005-0.05W(0)<0.
\end{equation}

To prove $W(0)>0$, set $\psi=\exp(-\Delta U/\D)>0$ and $W=\psi Z$.
With $\nu=0$, the water equation becomes
\begin{equation}\label{eq:positivewater}
 -(\D\psi Z_X)_X+(1+\beta U)\psi Z=\rho.
\end{equation}
The operator is coercive, so the Lax--Milgram theorem gives a unique periodic solution, and elliptic regularity makes it smooth.
The maximum principle gives $Z>0$ and hence $W>0$.
Therefore, $U_T<0$ at a point where $U=0$.
Any classical solution from these data becomes negative there for sufficiently small positive time.

\bibliographystyle{unsrt}
\begingroup
\small
\raggedright
\bibliography{references}
\endgroup
\end{document}